# Observation of the Purcell effect in high-index-contrast micropillars

A. J. Bennett, D. J. P. Ellis, A. J. Shields,

*Toshiba Research Europe Limited, Cambridge Research Laboratory, 260 Science Park, Milton Road, Cambridge, CB4 OWE, U. K.*

P. Atkinson,[a)] I. Farrer, and D. A. Ritchie

*Cavendish Laboratory, Cambridge University, JJ Thomson Avenue, Cambridge, CB3 0HE, U K.*

## Abstract

We have fabricated pillar microcavity samples with Bragg mirrors consisting of alternate layers of GaAs and Aluminium Oxide. Compared to the more widely studied GaAs/AlAs micropillars these mirrors can achieve higher reflectivities with fewer layer repeats and reduce the mode volume. We have studied a number of samples containing a low density of InGaAs/GaAs self assembled quantum dots in a cavity and here report observation of a three fold enhancement in the radiative lifetime of a quantum dot exciton state due to the Purcell effect.

**Main text**

Pillar-microcavities can increase the efficiency with which photons are collected and control the radiative dynamics of single-photon sources [1]. To date they have mainly been realised with Bragg mirrors consisting of $Al_xGa_{1-x}As$ layers of differing Al-content [2-5]. The small refractive index difference between GaAs ($n$ ~ 3.50) and AlAs ($n$ ~ 3.00) results in a reflectivity of only 0.6% at every interface so many layers are required to obtain up a high reflectivity. An alternative technique is to employ the wet-oxidation of AlAs [6] to form an oxide layer ("AlOx") with a much lower refractive index ($n$ ~ 1.53) than AlAs. A Bragg mirror consisting of alternate layers of GaAs and AlOx gives a reflectivity of ~ 15% at each interface. Consequently, several investigations have been carried out into the use of AlOx in Bragg mirrors and cavities [7-11], concentrating on vertical-emission light sources with relatively large lateral dimensions.

In this paper we highlight the advantages of using pillar microcavities with AlOx/GaAs Bragg mirrors as a single-photon source. We study the variation in cavity quality factor as a function of the pillar diameter and report that a Purcell effect can be observed in 2-3 μm diameter micropillars.

Figure 1 (a) shows the intensity transmission coefficient for both AlAs/GaAs and AlOx/GaAs mirrors as a function of the number of mirror repeats. For AlOx/GaAs cavities with 4 (7) periods above (below) the 1-wavelength thick GaAs spacer the planar cavity quality factor ($Q$) is ~15,000 (in the absence of absorption and optical imperfections), which is comparable to an AlAs/GaAs structure with 17 (25) periods. Figure 1 (d) and (e) show the magnitude of the electric field, $|E|$, calculated using an eigenmode solver model [12], along the axis of 2.0 μm diameter pillars with AlAs/GaAs or AlOx/GaAs mirrors. For the later case an increased reflectivity per interface reduces the volume of the mode, $V$, which is of particular relevance in determining the Purcell factor, $F_p$:

$$F_P = \frac{3Q}{4\pi^2 V}\left(\frac{\lambda}{n_{eff}}\right)^3 \qquad [1]$$

$V$ can be calculated by integrating the field intensity over a volume much larger than the mode and then normalizing to the maximum value field intensity, as stated in Eq. (2) [5].

$$V = \frac{\oint \varepsilon(\vec{r})\left|E(\vec{r})\right|^2 .dV}{\varepsilon_{\max}(\vec{r})\left|E_{\max}(\vec{r})\right|^2} \qquad [2]$$

In replacing the AlAs/GaAs Bragg mirrors with AlOx/GaAs the mode volume calculated for a 2 μm pillar is reduced from 19.5 to 3.2 $(\lambda/n_{eff})^3$. In general, $V/(\lambda/n_{eff})^3$ for AlOx/GaAs cavities is approximately one sixth of that in comparable AlAs/GaAs cavities for diameters between 1 and 10 μm. For example, in a 1.0 μm diameter AlOx/GaAs pillar $V/(\lambda/n_{eff})^3$ is only 1.6. However, it has been argued that the stronger spatial confinement of the mode increases photon loss in the transverse direction for pillars with diameters close to 1.0 μm [13], precluding the observation of strong coupling. Whether, this reduction in $Q$ can be ameliorated by modifying the spacer thickness [13] or the thicknesses of the first few mirror periods [14] remains to be investigated.

Before growth of the full structure oxidised test samples allowed us to measure a material contraction of 12% (using a surface profiler) and an AlOx refractive index of 1.53 (using ellipsometry), in agreement with existing literature [15]. Cavities were then grown with 4 (7) periods above (below) a 1-wavelength thick spacer. Despite the large thickness reduction the structural integrity of the oxidised pillars was good. After growth, processing was carried out with standard photolithography and a $SiCl_4$/Ar reactive ion etch followed by wet-oxidation at 400°C to form the AlOx layers.

Optical measurements were made at 4K in a continuous flow He cryostat using a spectrometer with a liquid nitrogen cooled charge-coupled device (CCD) or a Silicon avalanche-photodiode, for time resolved experiments. Reflectivity measurements were used to determine the quality factors of $HE_{11}$ modes in individual cavities using a white light source. An example of one such reflectivity spectrum is shown in the insert of Figure 2(b), recorded from a 2.5 μm diameter pillar. Similar measurements were taken for a number of pillars of each nominal size (Figure 2). In all pillars the $Q$ values measured are well below the predicted $Q$ of the planar cavity which is usually attributed to scattering from the sidewalls of the pillar. Figure 2(c) shows there is considerable variation in $Q$ for pillars of the same diameter, several times larger than is observed for the $Al_{0.98}Ga_{0.02}As$/GaAs pillars of comparable size [4, 5]. We also observe a large variation in mode wavelength between nominally identical pillars. This scatter is so large it is not possible to discern any trend in a plot of $Q$ versus the $HE_{11}$ wavelength (Figure 2(c)). This must be due to variations in layer thicknesses and diameter between pillars.

Firstly, the scatter in mode wavelength in Figure 2 could be due to variations in thickness of the 1-wavelength thick spacer layer between the Bragg mirrors. This could be the result of few-monolayer thickness variations due to “terracing” during sample growth. Thickness variations within each layer of the Bragg mirrors will also occur but these will be random and to some extent will average out. Assuming that the mirrors are fixed, the resonant mode wavelength shifts linearly by + 10.3nm for every nm of variation in the spacer layer in the AlOx/GaAs structure. This implies that the scatter in mode wavelength we observe can be explained by thickness variations of 10 monolayers, which is not unreasonable. For comparison in the AlAs/GaAs pillars the comparable shift is 4.1 nm, so this effect would be less pronounced. Variations in the thicknesses of the layers within each pillar will also reduce the $Q$ value [10].

Secondly, variations in diameter of nominally identical pillars will occur due to processing imperfections. A simple "waves-in-a-box" model shows that the wavelength of the $HE_{11}$ mode varies as $\lambda(r) = \lambda_0 - A/(n_{eff}r)^2$, where $\lambda_0$ is the planar cavity wavelength and $A$ a constant. This approximation agrees with the mode positions predicted by the full 3D model (Fig.1 (b)) for cavities with AlAs/GaAs and AlOx/GaAs mirrors. Thus, the reduced $n_{eff}$ in the AlOx/GaAs pillars leads to their $HE_{11}$ mode wavelength being almost twice as susceptible to variations in pillar diameter as in the all-semiconductor pillars.

Although 3.0 μm diameter pillars have variable mode wavelengths and $Q$, it was possible to identify a number of pillars with quantum dot exciton ($X$) emission lines spectrally close to the $HE_{11}$ mode. As these pillars have a diameter several times the mode wavelength we expect a well collimated emission pattern from the $HE_{11}$ mode that can be efficiently collected by a lens. Several micropillars were identified with similar Purcell factors, one of which we present data from in Figure 3.

The pillar has a diameter of 3.0 μm with the $HE_{11}$ at 910 nm and a $Q$ of 1500 from which we might expect a greater than one order of magnitude enhancement in the emission decay rate. Exciton ($X$) and biexciton ($X_2$) transitions can be identified by their linear and quadratic intensity variation with excitation power. At 3.6 K the $X$ line is detuned from the resonant mode by $\lambda_X - \lambda_{HE_{11}} = -0.93$ nm and the $X_2$ line is detuned by +0.45 nm. As the temperature is increased to 40 K the $X_2$ line detuning increases, its intensity falls rapidly and its radiative lifetime increases. In contrast, as the $X$ line is tuned onto resonance with the $HE_{11}$ mode at 40 K its intensity is increased by a factor of 30. The observed radiative lifetime corresponds to a Purcell factor of ~ 3. The discrepancy between the measured and predicted Purcell effect could be due to non-optimal spatial alignment of the QD with the cavity mode, for instance if the QD was not in the center of the pillar. However, in the course of this study a large number of pillars with $X$ states spectrally close to the cavity $HE_{11}$ mode were studied and none had Purcell factors larger than this. This suggests that our estimate of the mode volume is incorrect. If layers in the mirror deviate from one-quarter wave thickness the mode will penetrate further into the Bragg mirrors, increasing the mode volume. In addition scattering at pillar side walls may also affect the mode volume.

In conclusion, we have studied pillar microcavity single-photon sources where light is confined by high-index contrast Bragg mirrors. Experimentally, we observed Purcell factors of up to three in these structures. Large variations in quality factors and mode wavelengths between nominally identical pillars result from susceptibility of these structures to variations in pillar diameter and/or spacer layer thickness. Future improvements could be made through refining our processing or by utilising growth techniques which can reduce variations in layer thickness across the sample [16, 17].

## Figure Captions

Figure 1: Characteristics of Bragg mirrors and cavities formed from (open symbols) AlAs/GaAs and (filled symbols) AlOx/GaAs. (a) Intensity transmission coefficients for Bragg mirrors (light incident from GaAs, exiting into GaAs). (b) Wavelength of $HE_{11}$ mode as a function of pillar diameter. (c) Reflectivity spectra of (dotted line) 17/25 period AlAs/GaAs planar microcavity and (solid line) a 4/6 period AlOx/GaAs planar microcavity. (d) and (e) show the absolute electric field (line) and refractive index (grey filled plot) along the z-axis of 2.0 μm diameter micropillars with AlAs/GaAs mirrors and AlOx/GaAs mirrors, respectively.

Figure 2: (a) Experimental quality factors versus wavelength for $HE_{11}$ modes in a sample of AlOx/GaAs pillar microcavities at 5K. (b) Shows an example reflectivity measurement for a 2.5 μm diameter pillar (the circled data point in (a)) along with a lorentzian fit for a $Q$ of 2000. (c) Quality factors versus nominal diameter for AlOx/GaAs cavities.

Figure 3:(a) Wavelength of the $HE_{11}$ mode, $X_2$ and $X$ state in a 3.0 μm diameter pillar as a function of temperature. Shown in (b) is the radiative lifetime of the $X$ state as a function of the detuning between the $HE_{11}$ mode and the $X$ state. (c) shows two emission spectra at 3.6 and 40.0K.

**Author Footnotes**

[a)] now at: Max-Planck-Institut für Festkörperforschung, Heisenbergstr.1, D-70569 Stuttgart, Germany

**Figure 1**

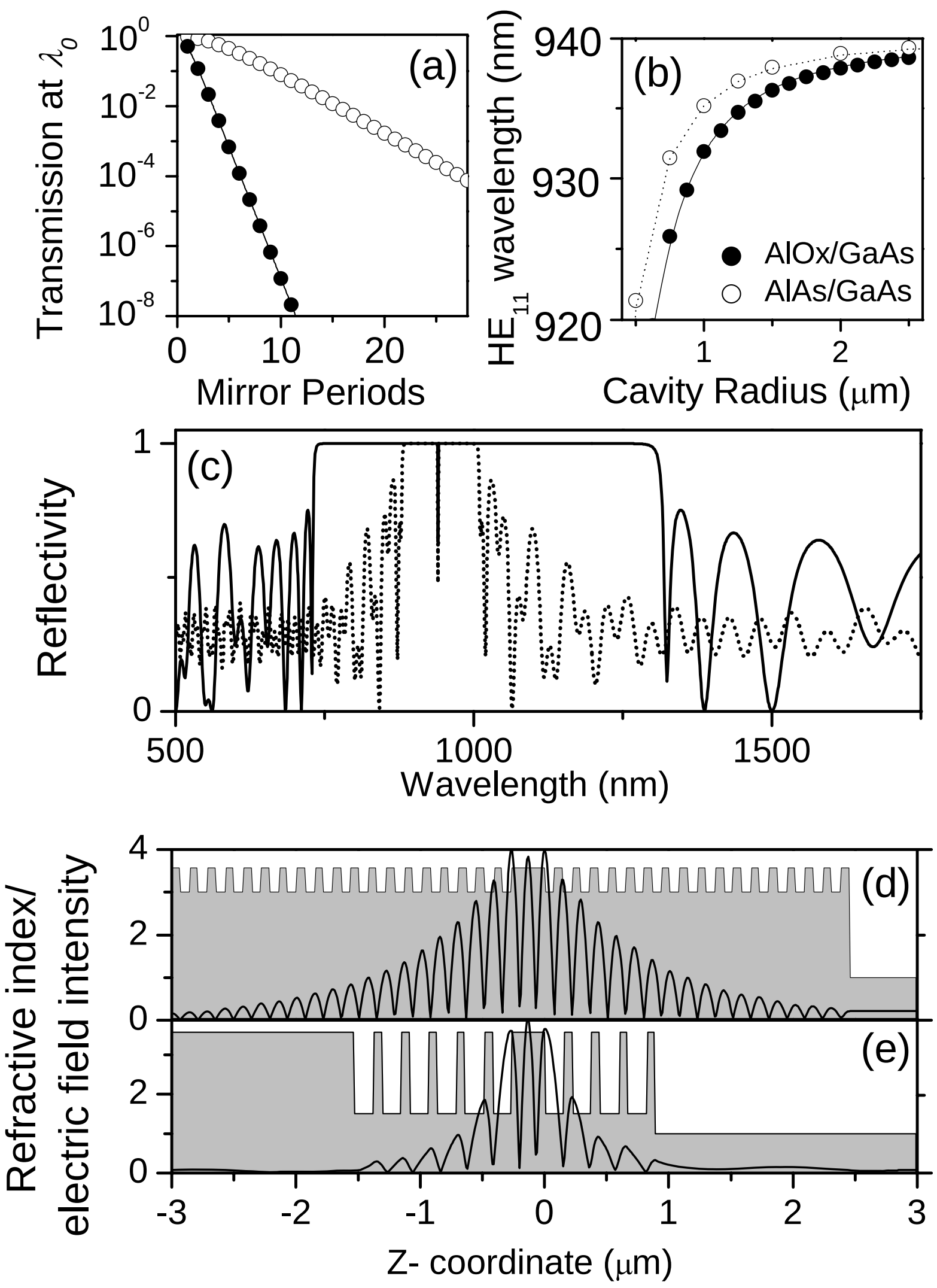

**Figure 2**

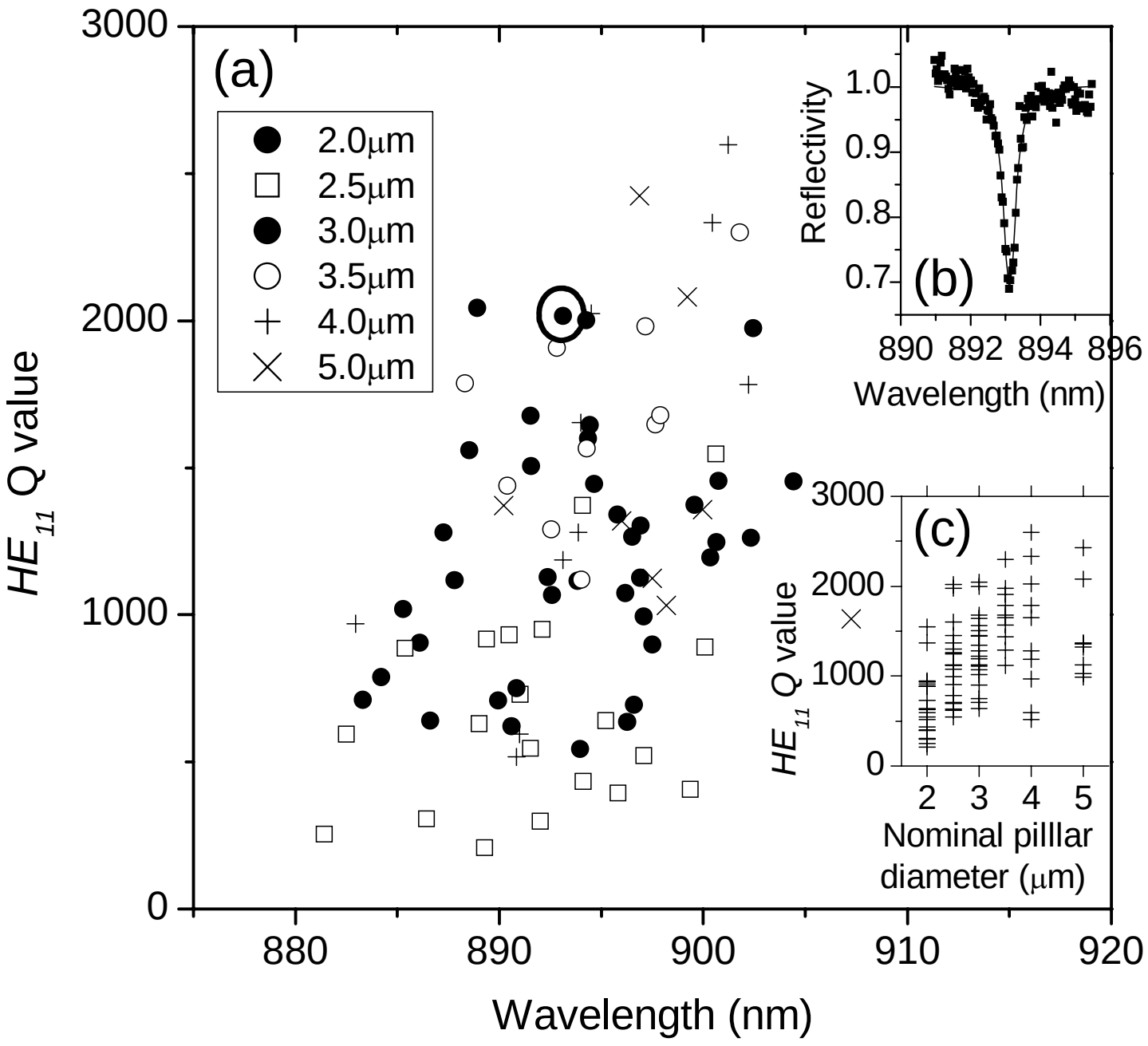

**Figure 3**

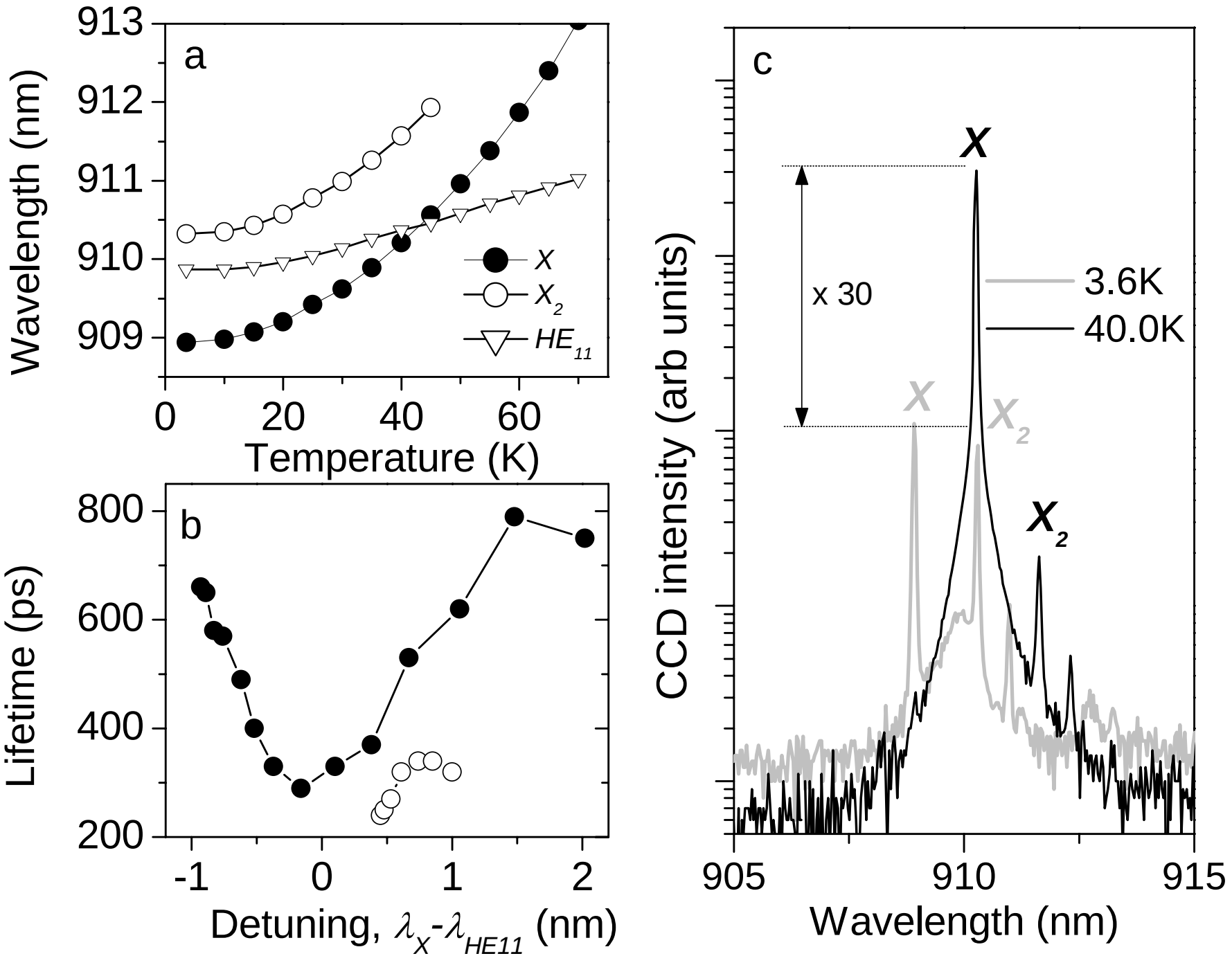